\documentclass[twocolumn,english,showpacs]{revtex4}
\usepackage[T1]{fontenc}
\usepackage[latin1]{inputenc}
\usepackage{amsmath}
\usepackage{graphicx}
\usepackage{amssymb}

\makeatletter

\usepackage{babel}
\makeatother
\begin{document}

\title{Finite temperature properties of the two-dimensional SU(2)
Kondo-necklace}

\author{Wolfram Brenig}

\affiliation{Technische Universit\"at Braunschweig, Mendelssohnstr. 3, 
38106 Braunschweig, Germany}

\date{\today}

\begin{abstract}
We analyse several thermodynamic properties of the two-dimensional Kondo
necklace using finite-temperature stochastic series expansion.  In
agreement with previous zero-temperature findings the model is shown to
exhibit a quantum critical point (QCP), separating an antiferromagnetic from
a paramagnetic dimerized state at a critical Kondo exchange-coupling
strength $J_{c}\approx 1.4$. We evaluate the temperature dependent uniform
and staggered structure factors as well as the uniform and staggered
susceptibilities and the local 'impurity' susceptibility close to the QCP
as well as in the ordered and quantum disordered phase. The crossover
between the classical, renormalized classical, and quantum critical regime
is analyzed as a function of temperature and Kondo coupling.
\end{abstract}


\maketitle
There is growing evidence that unconventional finite temperature properties
of many novel materials stem from zero-temperature phase transitions,
i.e. changes of the ground state symmetry as a function of some control
parameter. Prominent potential candidates to show such quantum phase
transitions are the cuprate superconductors \cite{Sachdev2003a,Alff2003a},
quantum magnets \cite{Bitko1996,Ruegg2003a}, and heavy-fermion or dense Kondo
systems \cite{Schroder2000a,Si2001a}. In the latter, quantum critical
points (QCPs) can arise from the competition between magnetic long-rage
ordered (LRO) and renormalized paramagnetic metallic or semimetallic
phases resulting from local Kondo-screening. This has been conjectured
early on by Doniach \cite{Doniach1977a}. Semimetallic behavior in
nonmagnetic states of Kondo lattice materials is typical for stoichiometric
{}``Kondo insulators'' like CeNi$_{1-x}$Pt$_{x}$Sn which undergoes
a para-to-antiferromagnetic transition at $x\approx 0.2...0.3$ 
\cite{Nishigori1993a,Kyogaku1992a}. A model for such materials is
 Kondo-Hubbard lattice
model (KHLM)\begin{equation}
H_{KH}=-t\sum _{lm,\sigma }c_{l\sigma }^{\dagger }c_{m\sigma }+
U\sum _{l}n_{l\uparrow }n_{l\downarrow }+
J\sum _{l,\alpha \beta }\mathbf{S}_{Pl}\cdot 
\mathbf{S}_{Il}\label{eq1}\end{equation}
with conduction electrons $c_{l\sigma }^{(\dagger )}$ of spin 
$\mathbf{S}_{Pl}$,
which are correlated via and on-site Coulomb repulsion $U$, and coupled
by antiferromagnetic (AFM) Kondo-exchange to localized spins $\mathbf{S}_{Il}$
at sites $l$. At half filling on bipartite lattices in D$\geq 2$
dimensions and in the strong-coupling limit $U/t\gg 1$ the HKLM shows
AFM LRO if the conduction-electron superexchange $j\sim t^{2}/U$
dominates the Kondo scale $T_{K}\sim t\, \exp \left(-1/\rho J\right)$
where $\rho $ refers to the DOS \cite{CommentRKKY}. Kondo screening
will prevail if $T_{K}/j\gg 1$. On 2D square lattices the critical
coupling $j_{c}\left(U/t\right)$ has been determined at temperature
$T=0$ using projector QMC \cite{Feldbacher2001a} and bond-operator
methods \cite{Jurecka2001a}. In the strong coupling limit and at
half filling Eqn. (\ref{eq1}) simplifies to the $SU\left(2\right)$-symmetric
so-called Kondo-necklace (SKN)\begin{equation}
H_{SKN}=j\sum _{lm}\mathbf{S}_{Pl}\cdot \mathbf{S}_{Pm}+J\sum _{l}
\mathbf{S}_{Pl}\cdot S_{Il}\label{eq2}\end{equation}
with $j\equiv 1$ hereafter. In this work we will focus on the 2D
square lattice, where at $T=0$ the SKN has been investigated by bond-operator
methods, series expansion and exact diagonalization
\cite{Matsushita1997a,Kotov1998a,ZhangComment2}.
These studies located  a QCP at $J_{c}\sim 1.370\ldots 1.408$
separating AFM LRO from a gapped spin-dimer phase. The latter can
be viewed as the strong-coupling analog of the Kondo-screened paramagnetic
state of the KHLM.

While the ground state properties of the 2D SKN have been studied
rather extensively, thermodynamic and finite temperature critical
properties of the SKN remain an open issue. Therefore the aim of this
work is to shed light on the 2D SKN at finite temperatures using a
quantum Monte-Carlo (QMC) approach. To this end we employ the stochastic
series expansion (SSE) with loop-updates introduced by Sandvik and
Syljuasen in Refs. \cite{Sandvik1999a,Syljuasen2002a} to which we
refer the reader for details on this approach.

\begin{figure}[tb]
\begin{center}\includegraphics[  width=0.95\columnwidth,
  keepaspectratio]{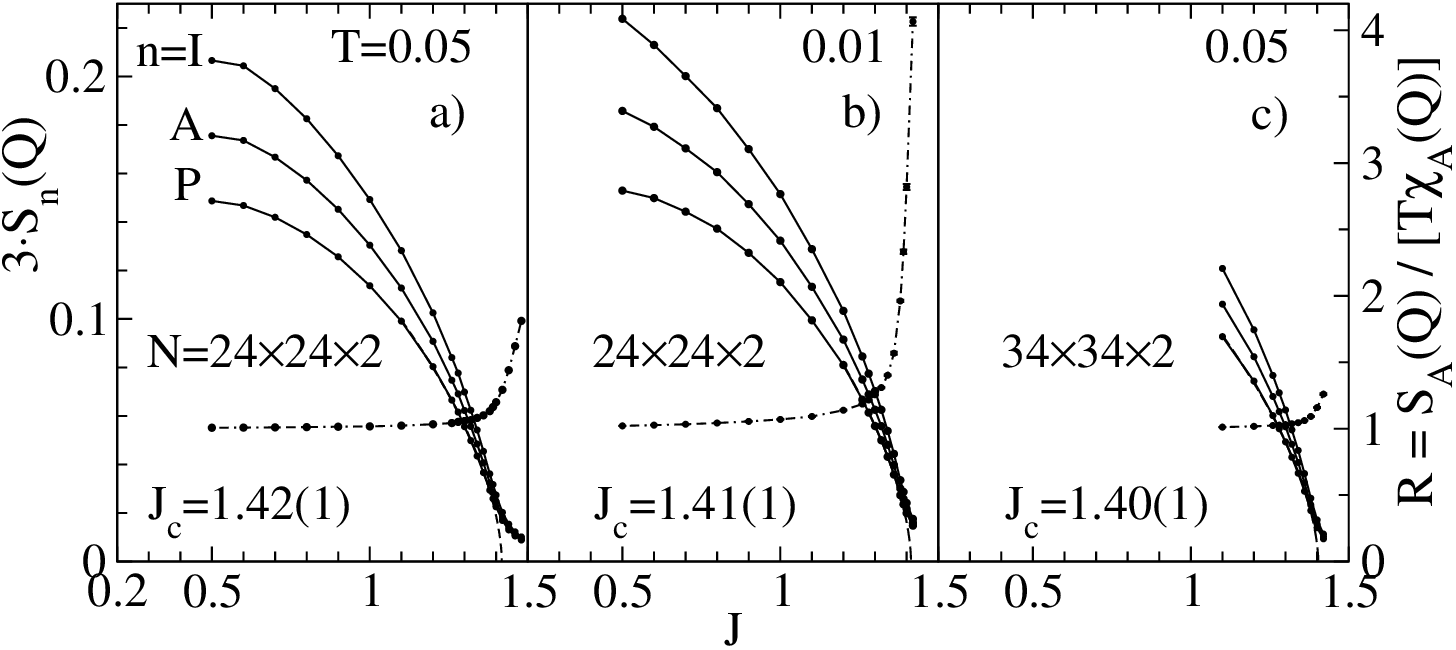}\end{center}
\vspace{-5mm}

\caption{\label{fig1} Solid: staggered structure factor $S_{n}\left(
\mathbf{Q}\right)$
vs. $J$ close to the QCP at low-temperatures and for two system sizes:
a) $T=0.05$, $L=24$, b) $T=0.01$, $L=24$, and c) $T=0.05$, $L=34$.
Dashed: fits of $S_{P}\left(\mathbf{Q}\right)$ to $c\left[J_{c}-
J\right]^{\nu }$
for $J\gtrsim 1$ with $J_{c}$ indicated per panel. The fits depart
visibly from $S_{P}\left(\mathbf{Q}\right)$ only close to $J\approx 1.4$.
Dashed-dotted: ratio of total staggered structure factor to susceptibility
times $T$. If not indicated, statistical errors are less than the
solid-circle marker size.}
\end{figure}

We start by discussing the longitudinal staggered structure factor
\begin{equation}
S_{n}\left(\mathbf{Q}\right)=\left\langle \left(m_{n\mathbf{Q}}^{z}
\right)^{2}\right\rangle \, \, ,\label{eq3}\end{equation}
 where $m_{n\mathbf{Q}}^{z}=\sum _{l}S_{nl}^{z}\exp \left(i\mathbf{Q}\cdot 
\mathbf{r}_{l}\right)/N_{n}$
is the staggered magnetization with $\mathbf{Q}=\left(\pi ,\pi ,\pi \right)$.
$m_{n\mathbf{Q}}^{z}$ selects between $n=P,I,A$, for which $\mathbf{r}_{l}$
runs over the 'conduction electron plane' for $n=P$, the 'Kondo sites'
sites for $n=I$, and all sites for $n=A$. Fig. \ref{fig1} shows
the squared staggered moment $M_{\mathbf{Q}}^{2}=3S_{n}\left(\mathbf{Q}
\right)$
vs. $J$ at low temperatures. The system sizes $N_{A}$ are $L\times L\times 
2\equiv N$
with periodic boundary conditions (PBC) in the planar directions and
$L=24$ and $34$. In all three panels $M_{\mathbf{Q}}^{2}$ is finite
below a critical value of $J=J_{c}$ and drops to approximately zero
for $J>J_{c}$. We identify $J_{c}$ with the QCP and expect AFM LRO
for $J<J_{c}$ in the thermodynamic limit at $T=0$. For $J>J_{c}$
we find no other transitions, i.e. the systems connects adiabatically
to the limit of $J=\infty $. Therefore it is dimerized. At fixed
$N$, $M_{Q}^{2}$ will saturate for $T\rightarrow 0$ due to finite
size gaps. For $L=24$ this is the case in Fig. \ref{fig1} b) for
$J\gtrsim 1$. Fig. \ref{fig1} allows no conclusion about the magnitude
of the $T=0$ order parameter, which requires finite size scaling
analysis \cite{Brenig2005a}. The critical coupling however can be
extracted efficiently from these results since $J_{c}$ is almost
invariant to increasing $N$ or lowering $T$ relative to the parameters
in Fig. \ref{fig1}. To determine $J_{c}$ we fit $M_{\mathbf{Q}}^{2}$
to a power law $M_{\mathbf{Q}}^{2}\approx c\left[J_{c}-J\right]^{\nu }$
in its region of negative curvature and for $J\gtrsim 1$. This procedure
depends only little on the interval of $J$ fitted to. The resulting
scatter of $J_{c}$ is taken to be a measure of the error and is displayed
in Fig. \ref{fig1} a)-c). We find that $J_{c}\approx 1.41(2)$. This
agrees with $J_{c}\approx 1.41$($1.39$) from $T=0$ series-expansion
\cite{Matsushita1997a}(\cite{Kotov1998a}) and is also close to 
$J_{c}\approx 1.37$
from bond-operator Brückner theory \cite{Kotov1998a,comment1}. A
critical value of $J_{c}\approx 1.4\left(4t^{2}/U\right)$ is also
consistent with projector-QMC at $T=0$ for the KHLM \cite{Feldbacher2001a}.

For small $J$, $S_{n}\left(\mathbf{Q}\right)$ remains strongly temperature
dependent down to $T\ll 1$ which is due to the near decoupling of
the $I$-sites from the planar sites leading to a Curie-like contribution
which is cutoff only at very low $T$. This is visible already at
$J\lesssim 0.6$, by comparing $S_{I}\left(Q\right)$ in panels a)
and b) of Fig. \ref{fig1}. 

\begin{figure}[tb]
\begin{center}\includegraphics[  width=0.95\columnwidth,
  keepaspectratio]{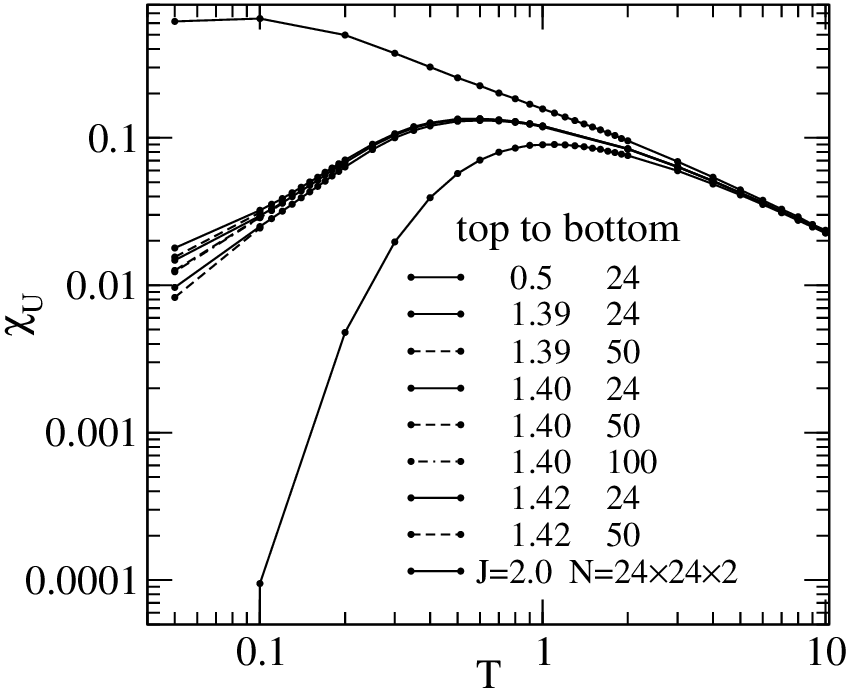}\end{center}
\vspace{-5mm}

\caption{\label{fig2} Uniform susceptibility $\chi _{u}$ vs.
temperature for $0.05\leq T\leq 10$ and $0.5\leq J\leq 2.0$
at $L=$24 (solid) and  for $1.39\leq J\leq 1.42$ at
$L=50\left(\textrm{100}\right)$ (dashed(dashed-dotted)).
Statistical errors are less than the solid-circle marker size.
The difference between $L=$50 and 100 at $J=1.40$ remains below
the statistical error for all T depicted. Legends label plots
from top to bottom}\vspace{-5mm}

\end{figure}

In addition to $S_{n}\left(\mathbf{Q}\right)$ Fig. \ref{fig1} includes
results for the longitudinal staggered susceptibility\begin{equation}
\chi _{n}\left(\mathbf{Q}\right)=\int _{0}^{\beta }d\tau \left\langle 
m_{n\mathbf{Q}}^{z}\left(\tau \right)m_{n\mathbf{Q}}^{z}\right\rangle 
\label{eq4}\end{equation}
 for $n=A$ which have been encapsulated in the ratio\begin{equation}
R=\frac{S_{A}\left(\mathbf{Q}\right)}{T\chi _{A}\left(\mathbf{Q}\right)}\, 
.\label{eq5}\end{equation}
 This ratio relates the analysis of $\chi _{A}\left(\mathbf{Q}\right)$
to that of the AFM non-linear $\sigma $-model (NL$\sigma $M) 
\cite{Chakravarty1989a,Chubukov1994a,Sokol1994a}.
From there it is expected that in the classical high-$T$, as well
as in the low-$T$ renormalized classical regime $R=1$, while $R\neq 1$
in the quantum critical regime. While this is consistent with $R\left(
J\right)$
in Fig. \ref{fig1}, we will clarify later that the deviations of
$R$ from unity for $J\approx J_{c}$ are strongly affected by finite
size effects.

\begin{figure}[tb]
\begin{center}\includegraphics[  width=0.95\columnwidth,
  keepaspectratio]{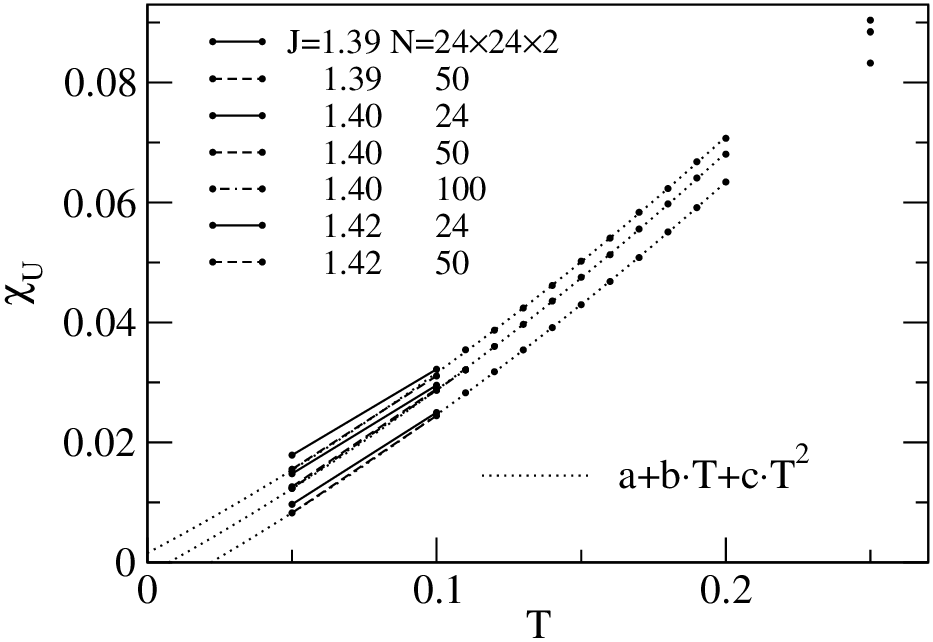}\end{center}
\vspace{-5mm}

\caption{\label{fig3}
Low-temperature uniform susceptibility $\chi _{u}\left(T\right)$
close to the QCP for $J=1.39,\, 1.40$, and $1.42$ (from top to bottom).
Dotted curves: fits of $\chi _{u}\left(T\right)$ to $a+bT+cT^2$.
System sizes $L=$24 (solid), 50 (dashed), 100 (dashed-dotted) are
indicated for $T\lesssim$0.1. Statistical errors are less than
the solid circle marker size. For $T\geq 0.11$ finite size effects
are below statistical error and results for $J=1.40$(1.39, 1.42)
refer to $L=100$(24) only. The difference between $L=$50 and 100
at $J=1.40$ remains below the statistical error for all T depicted.
 }\vspace{-5mm}

\end{figure}

Next we discuss the uniform susceptibility\begin{equation}
\chi _{u}=\beta \left\langle m^{2}\right\rangle \, \, ,\label{eq6}
\end{equation}
 where $m=\sum _{l}S_{Al}^{z}/N_{A}$ is the total magnetization.
In contrast to Eqn. (\ref{eq4}), the uniform susceptibility reduces
to a simple expectation value, since $\left[H_{SKN},m\right]=0$.
Fig. \ref{eq2} is a log-log plot of the dependence of $\chi _{u}$
on temperature over more than two decades $0.05\leq T\leq 10$ and
for $0.5\leq J\leq 2$ with system sizes $L=24$, 50 and 100. For $T\geq 0.1$
finite-size effects are negligible if $L\geq 24$. For $0.05\leq T\leq 0.1$
finite-size effects, albeit small, have been considered for $J$ in
the vicinity of the QCP. As can be seen from the near identity of
results with $L=50$ and 100 at $J=1.40$ in Fig. \ref{fig2}, it is sufficient
to choose $L\gtrsim 50$ to reach the thermodynamic limit for all
temperatures studied. For $T\gtrsim 2$ the uniform susceptibility turns
Curie-like, independent of $J$. For $J>J_{c}$ the spin-spectrum
develops a gap $\Delta $ which implies a low-temperature behavior
$\chi _{u}\propto \exp \left(-\beta \Delta \right)$. This is consistent
with Fig. \ref{eq2}, where to within statistical error
$\chi _{u}\left(T=0.05,J=2\right)=0$.
For $0<J<J_{c}$ AFM LRO occurs at $T=0$, which agrees with the saturation
of $\chi _{u}\left(T\rightarrow 0\right)=\chi _{u}^{0}$ shown in
the figure. We note, that as $J$ vanishes $\chi _{u}^{0}$ will diverge
due to the Curie contribution form the impurity spins.

\begin{figure}[tb]
\begin{center}\includegraphics[  width=0.95\columnwidth]{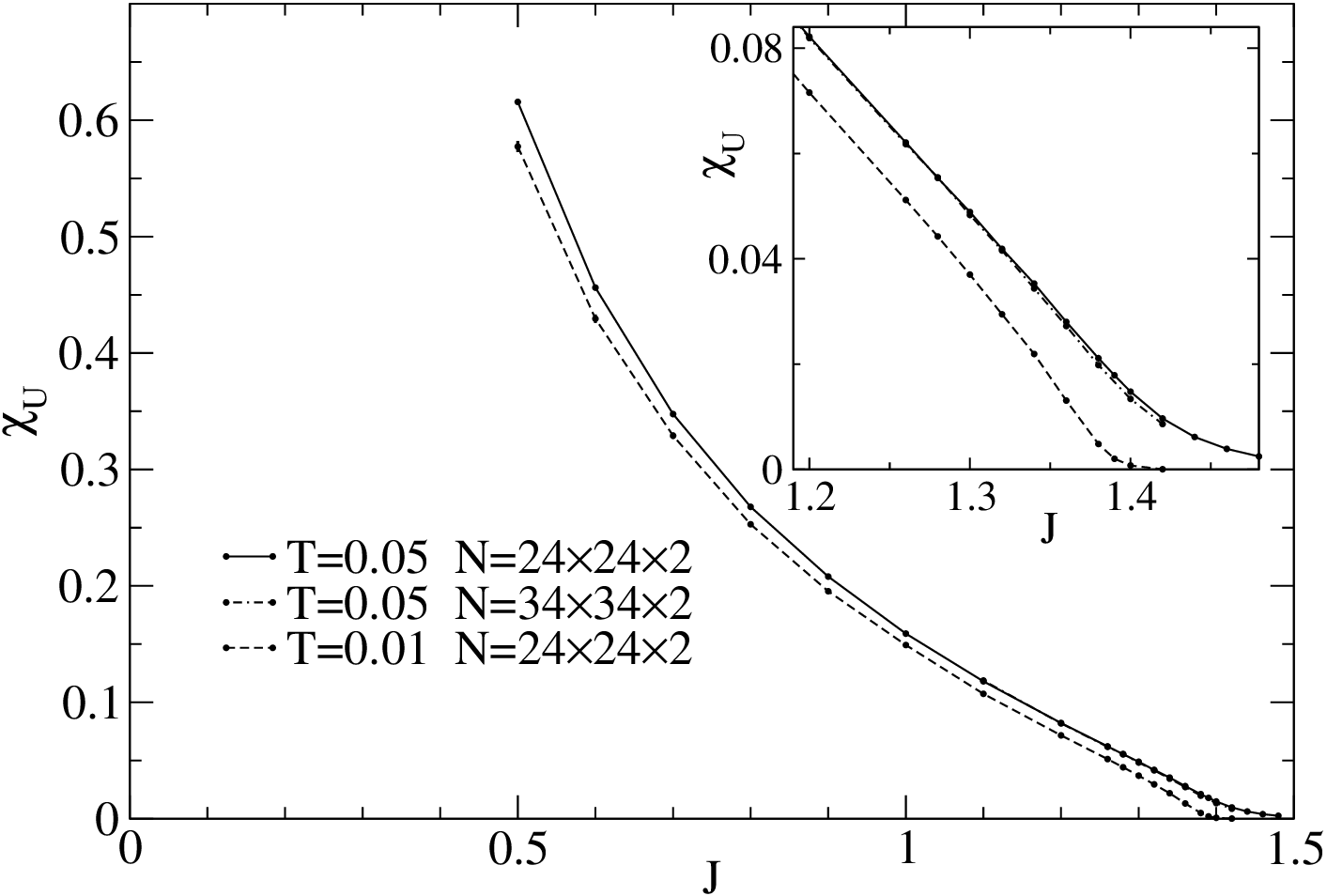}\end{center}
\vspace{-5mm}

\caption{\label{fig4} Uniform susceptibility $\chi _{u}$ vs. $J$ on approaching
the QCP for two temperatures $T=0.05\left(0.01\right)$ (solid(dashed)
line) at fixed $L=24$, and comparing two system sizes $L=24\left(34\right)$
(solid(dashed-dotted) line) at fixed $T=0.05$. Finite size effects
are exceedingly small and are shown in the inset for better visibility.
Statistical errors are less than the solid-circle marker size.}\vspace{-5mm}

\end{figure}

At the QCP we expect scaling of the uniform susceptibility. Indeed,
for $J\approx J_{c}$, and at low temperatures $\chi _{u}$
follows nearly straight lines in Fig. \ref{fig2}. A close-up
of this low-$T$ region, shown in Fig. \ref{fig3}, evidences a weak
curvature of $\chi_u(T)$ independent of the system size. These
results allow excellent fits to a scale-free behavior of the
form $\chi _{u}\approx a+bT^{c}$ with $c\approx 1.25$. However,
this exponent differs from that obtained in the NL$\sigma$M, i.e.
$c=1$ \cite{Chakravarty1989a,Chubukov1994a}. Assuming the SKN to
be of the same universality class than the NL$\sigma$M we are forced
to treat the curvature in Fig. \ref{fig3} as deviations from
scaling present already at rather low temperature. As is shown
in Fig. \ref{fig3}, a reasonable description of the QMC results
can be obtained including a second order nonuniversal contribution.
This behavior should be contrasted against
the AFM bilayer Heisenberg model, where critical linear $T$-scaling
has been found in a comparable temperature range \cite{Shevchenko2000a}. 
At the QCP $\chi_u$ vanishes for $T\rightarrow 0$ due to the opening
of the spin gap and for $J>J_c$ exponential behavior should replace
the scaling. Vanishing of the offset $a$ at
$J_{c}=1.40(1)$ in Fig. \ref{fig3} is consistent with  as from the static
structure factor.

For the sake of consistency it is interesting to consider the uniform
susceptibility $\chi _{u}$ also as a function of $J$ in the low-temperature
limit $T\ll J$. On approaching the QCP from the LRO side one expects
$\chi _{u}$ to vanish due to the incipient spin gap. The corresponding
QMC results are shown in \ref{fig4}. Extracting $J_{c}$ from this
figure is less straightforward, both, due to the sizeable temperature
variation and to the lack of a scaling prescription for $\chi _{u}$
as $J\rightarrow J_{c}$. Nevertheless, as can be seen in the inset,
a value of $J\approx 1.4$ for the QCP is consistent with the suppression
of $\chi _{u}$.

\begin{figure}[t]
\begin{center}\includegraphics[  width=0.95\columnwidth]{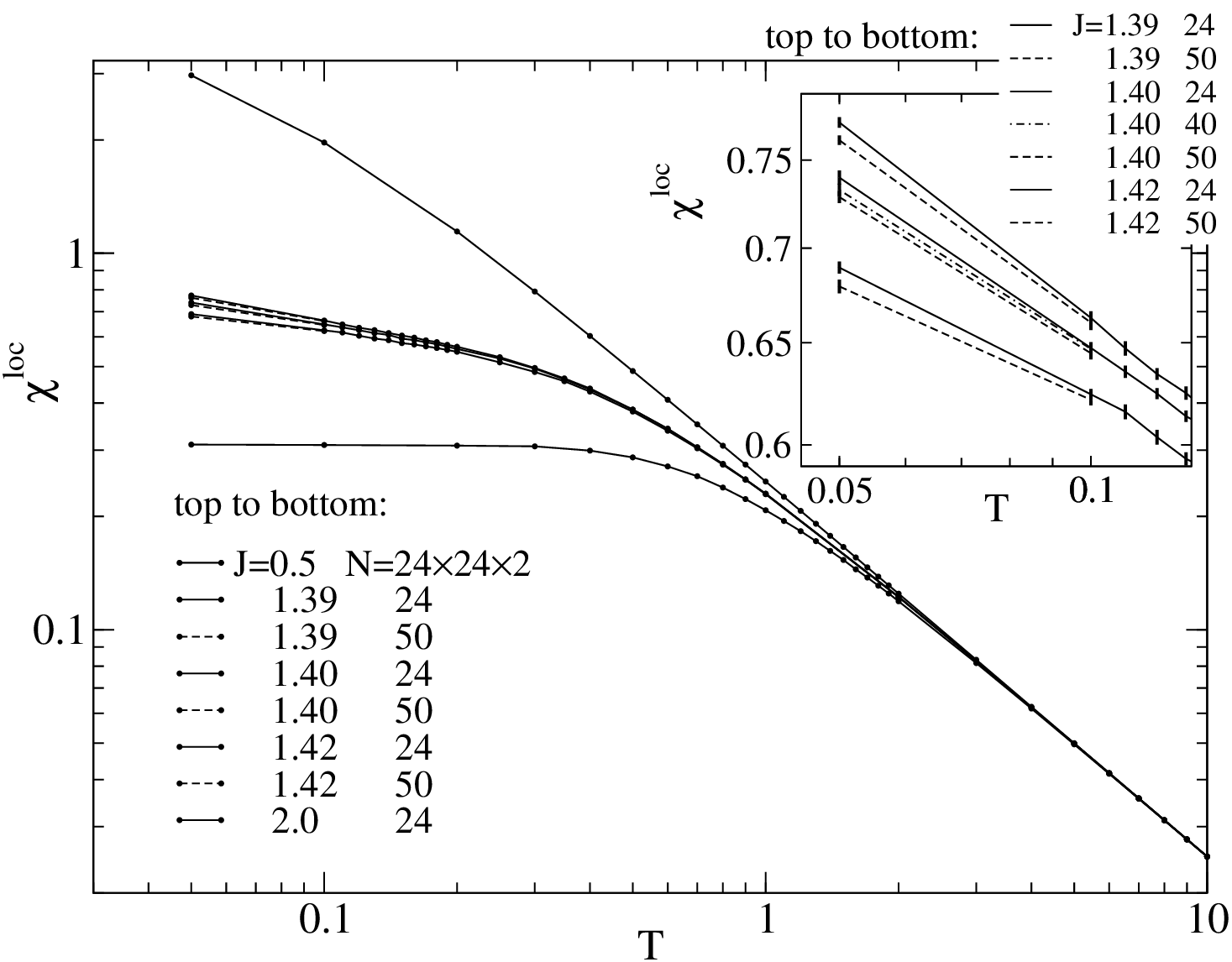}\end{center}
\vspace{-5mm}

\caption{\label{fig5} Impurity susceptibility $\chi ^{loc}$ vs. temperature.
Solid line with solid circle markers: $L=24$ for $0.05\leq T\leq 10$
and $0.5\leq J\leq 2.0$. Dashed line with solid circle markers: $L=50$
for $0.05\leq T\leq 0.1$ and $1.39\leq J\leq 1.42$. Inset: low-$T$
region including additional results for $L=40$ at $J=1.40$ (Dashed-dotted).
Legends refer to lines from top to bottom. Statistical errors are
less than the solid-circle marker size in the main panel and
are indicated by bars in the inset. In the quantum critical regime
$\chi ^{loc}$ displays a cross-over region with
power-law behavior $\chi ^{loc}
\approx cT^{-\alpha }$ with $\alpha \approx 0.20(5)$.
}\vspace{-5mm}

\end{figure}

Now we turn to the individual impurity-spins longitudinal susceptibility
\begin{equation}
\chi ^{loc}=\int _{0}^{\beta }d\tau \left\langle T_{\tau
 }S_{lI}^{z}\left(\tau
 \right)S_{lI}^{z}\right\rangle \label{eq7}\end{equation}
where $l$ refers to a particular site, say $l=0$ within the 'Kondo
spin' layer $I$. Fig. \ref{fig5} shows a log-log plot of $\chi ^{loc}$
vs. $T$ for $0.05\leq T\leq 10$ and $0.5\leq J\leq 2$. At $J=0$
$\chi ^{loc}$ obeys Curie's law. For $J\neq 0$ but $J<J_{c}$, we
expect $\chi ^{loc}$ to saturate at some cross-over temperature $T^{\star }
\lesssim J$
due to the coupling of the impurity spin to the planar moments within
the AFM LRO state. In agreement with this, Fig. \ref{fig5} signals
a departure from $\chi ^{loc}\propto T^{-1}$ for $T\approx 0.2$
at $J=0.5$, i.e. in the AFM LRO state. Similarly, for $J>J_{c}$
we expect a Pauli-like saturation of $\chi ^{loc}$ for $T\leq T^{\star }$
with $T^{\star }\lesssim J$ due to the local dimer formation between
the impurity spins and the planar sites. This can also be seen in
Fig. \ref{fig5} for $J=2$. The interesting point of Fig. \ref{fig5}
however, is that it suggests a cross-over from the high-temperature
Curie behavior to a region of power-law behavior $\chi ^{loc}\propto 
T^{-\alpha }$
with an exponent $\alpha $ different from unity in the vicinity of
the QCP. Future QMC analysis should focus on additional data in the
thermodynamic limit at $T<0.05$ to elaborate on this observation.
From Fig. \ref{fig5} we extract $\alpha \approx 0.20(5)$ at $J=1.40$.
The error on this exponent is rather large, due to the error in determining
the QCP and due to the temperature range of only one decade to fit
to. Regarding finite-site effects, the situation for $\chi ^{loc}$
is similar to that for $\chi _{u}$. As shown in the inset of Fig.
\ref{fig5}, in the vicinity of the QCP the thermodynamic limit is
reached for $L\gtrsim 24(50)$ if $T\gtrsim 0.1(0.05)$. Similar effects
are expected at the lowest temperature $T=0.05$ for $J=0.5$ and
$2.0$ and have not been considered.

\begin{figure}[tb]
\begin{center}\includegraphics[  width=0.95\columnwidth]{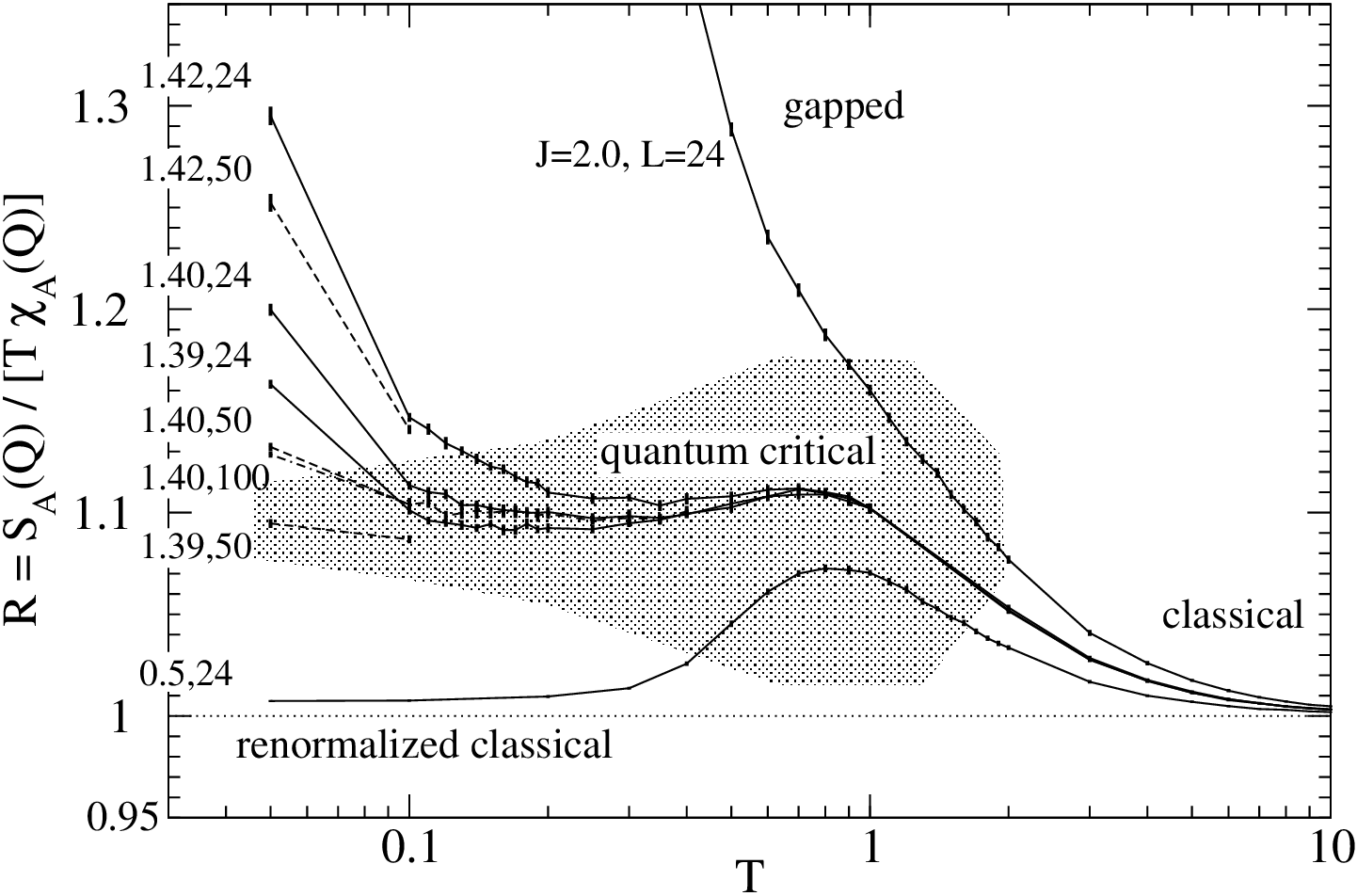}\end{center}
\vspace{-5mm}

\caption{\label{fig6}Solid(dashed){[}dashed-dotted{]}: ratio $R$ of the
total staggered structure factor and susceptibility vs. temperature
for $0.5\leq J\leq 2.0$ and $L=24$($50$){[}$100${]}. In the classical
(renormalized classical) regime, i.e. $T\rightarrow \infty $ 
($T\rightarrow 0,\, J<J_{c}$),
$R$ is expected to be $1$. In the gapped state $R1/T$ as $T\rightarrow 0$.
In the quantum critical regime, roughly sketched by the shaded region,
$R$ differs from unity, however approaching a constant as $T\rightarrow 0$.
Size of statistical errors is given by vertical bars.}\vspace{-5mm}

\end{figure}

Finally we analyze the temperature dependence of the ratio of the
total staggered structure factor to the total staggered susceptibility
of Eqn. (\ref{eq5}). This is shown in Fig. \ref{fig6} for $0.05\leq T\leq 10$
and $0.5\leq J\leq 2$. Nonlinear error propagation of the QMC data
through Eqn. (\ref{eq5}) leads to substantially larger statistical
errors on $R$ as compared to the remaining quantities evaluated in
this work. Several properties of $R$ can be realized based on general
grounds. First, for $T\gg max\left\{ J,1\right\} $, i.e. in the \emph{classical
regime}, $\chi _{A}\left(\mathbf{q}\right)=S_{A}\left(\mathbf{q}\right)/T$
for any wave vector $\mathbf{q}$ and therefore $R\rightarrow 1$.
This behavior of $R$ is obeyed for all values of $J$ displayed
in Fig. \ref{fig6}. Next, we note that the zero-temperature limit
of the ratio $S_{A}\left(\mathbf{Q}\right)/\chi _{A}\left(\mathbf{Q}\right)$
will be a $T$-independent constant whenever the system has no LRO
at the wave vector $\mathbf{Q}$ and is gapped. This is true
for any finite system, where $S_{A}\left(\mathbf{Q}\right)$
and $\chi _{A}\left(\mathbf{Q}\right)$ will both saturate at finite
values as $T\rightarrow 0$. It is also true in the thermodynamic
limit where $S_{A}\left(\mathbf{Q}\right)$ and $\chi _{A}\left(
\mathbf{Q}\right)$
will both be exponentially activated to leading order. Therefore $R\propto 1/T$
in the thermodynamic limit in the \emph{quantum disordered} regime,
i.e. for $J>J_{c}$, which is consistent with the increase of $R$
in Fig. \ref{fig6} for $J=2$. In addition to this, $R\propto 1/T$
for any other value of $J$ below a characteristic temperature set
by finite size gaps. This is particularly obvious for $J\approx J_{c}$
where strong finite size effects occur for $T<0.1$. This
effect also sets the magnitude of $R$ for $J\approx J_{c}$ 
in Fig. \ref{fig1}.
Finally, in the AFM LRO phase the systems allows for a classical description
in terms of the order parameter modes leading to a \emph{renormalized
classical regime} for which $R\left(T\rightarrow 0\right)=1$ again
\cite{Chakravarty1989a,Chubukov1994a}. This is consistent with the
behavior for $J=0.5$ in Fig. \ref{fig6} and with $R$ in Fig. \ref{fig1}.
From a comparison of $R$ at fixed $T$ and identical $J$ in panel
a) and c) of the latter figure one can also deduce that the small
difference between $R$ and unity in the renormalized classical regime
decreases upon increase of $L$. Lowering the temperature from the
classical to the renormalized classical regime, one crosses the \emph{quantum
critical regime} in which $R>1$ due to quantum fluctuations 
\cite{Chubukov1994a,Sokol1994a}.
For $J<J_{c}$ this regime has a finite extend in temperature only.
Close to the QCP however, i.e. for $J=1.4$ and $1.39$, Fig. \ref{fig6}
strongly suggests that $R$ approaches a temperature independent constant
$R\approx 1.10(1)$ as $T\rightarrow 0$. Analysis of the NL$\sigma$M
has resulted in $R=1.09$ \cite{Chubukov1994a,Sokol1994a}. In turn the
quantum critical regime starts at $T\sim 1$ and extends down to $T=0$
at the QCP. Unfortunately, in
this regime, $R$ is is very sensitive to the system size. This will
be the issue of future QMC studies\cite{Brenig2005a}. 

\emph{Acknowledgements} Fruitful discussions with E. Dagotto and the
kind hospitality of the National High Magnetic Field laboratory at
Florida State University in the early stages of this project are gratefully
acknowledged. A comment of M. Vojta on the scaling analysis is appreciated.
Part of this work has been supported by DFG grant BR 1084/1-3.


\end{document}